\documentclass{ws-procs9x6}
 \usepackage{wrapfig}
\begin{document}

\title{Recent results from the Tevatron}

\author{C.~ROYON }

\address{DAPNIA/Service de physique des particules, CEA/Saclay, 91191 
Gif-sur-Yvette cedex, France}

\maketitle

\abstracts{
We give the most recent results from the D\O\ and CDF experiments at the
Tevatron. 
}

The luminosity accumulated by the D\O\ and CDF experiments and used for many
analyses shown in this report is of the order of 1
fb$^{-1}$ per experiment and the efficiency of data taking is above 90\% for both experiments.
We will describe in this paper some of the newest results obtained by both
collaborations especially on QCD, diffraction, electroweak, top, $b$ physics as
well as the search for the Higgs boson and supersymmetry.

\section{QCD results}

\begin{wrapfigure}{l}{4.5cm}
\vspace{-1cm}
\includegraphics[clip,width=4.5cm,height=4.5cm]{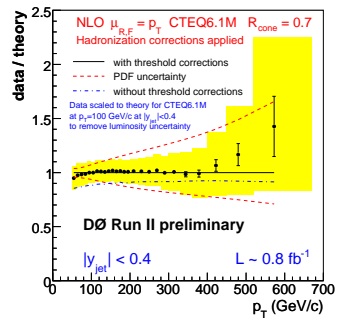}
\caption{Data/Theory for the jet inclusive $p_T$ cross section in the most
central bin in
rapidity using the 0.7 jet cone algorithm from the D\O\ experiment (see text).}
\end{wrapfigure}

\subsection{Measurement of the inclusive jet $p_T$ cross section}

Both CDF and D\O\ collaborations measured the jet inclusive $p_T$ cross section
using either the cone or $k_T$ jet algorithm. This measurement can constrain
further the gluon density at high $x$ - which is the one of the limiting
factors in some analyses beyond the standard model at the Tevatron and the
LHC like the search for the stop quark, and extra dimensions - and is also
sensitive to quark and gluon substructure. The D\O\ collaboration performed this
measurement in two jet rapidity bins ($|y| < 0.4$ and $0.4 < |y| < 0.8$) using a
luminosity of 0.8 fb$^{-1}$.  Data at particle level are compared to NLO calculations
including threshold corrections \cite{nlojet++,kidonakis} using the 
CTEQ6.1M parametrisation \cite{cteq6}, 
corrected for hadronisation effects \cite{Mikko}. Let us
note that the data have been scaled to NLO predictions for the $|y|<0.4$
bin at $p_T$=100 GeV since the evaluation of the luminosity of the full sample 
was in progress. The results are presented in Fig. 1 for the most central bin
in rapidity as the ratio data over
theory. The difference in behaviour at high $p_T$ between both bins
in rapidity is due mainly to
statistical effects in the jet energy scale determination \cite{Mikko}. Data are found to be
in good agreeement with NLO calculations.
The threshold correction effects
are represented by the difference between the blue dashed dotted line and the
black horizontal line and the uncertainty on PDFs by the red dashed line. The
present systematics displayed as the yellow band is of the same order of
magnitude as the uncertainty on PDFs.

\begin{wrapfigure}{l}{6.5cm}
\vspace{-0.5cm}
\includegraphics[clip,width=6.5cm,height=6.5cm]{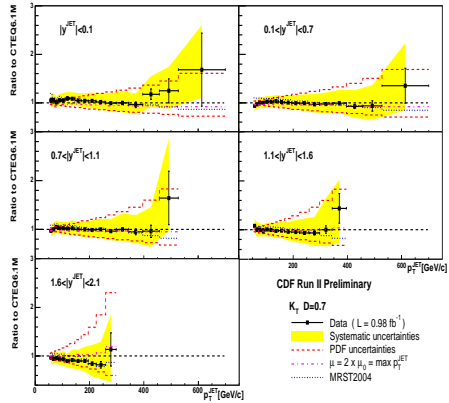}
\caption{Data/Theory for the jet inclusive $p_T$ cross section in five bins of
rapidity using the $k_T$ jet cone algorithm from the CDF experiment (see text).}
\end{wrapfigure}

The CDF collaboration also performed the
measurement of the jet inclusive $p_T$ spectrum using the 0.7 cone algorithm
for $0.1 < |y| <0.7$ at the parton level and the data are also found to be in
good agreement with NLO calculations \cite{cdfjets}.

%\begin{figure}[h]
%\centerline{\epsfxsize=3.5in\epsfbox{AllRatioCSoNLO.eps}}   
%\caption{Data/Theory for the jet inclusive $p_T$ cross section in five bins of
%rapidity using the $k_T$ jet cone algorithm from the CDF experiment (see text). }
%\end{figure}

The CDF collaboration also measured the inclusive jet $p_T$ spectrum using the
$k_T$ algorithm with about 1 fb$^{-1}$ and the results are displayed in 
Fig. 2 in five bins of
rapidity \cite{cdfjets} .  There is a good agreement between theory and experiment within
systematics (displayed as a yellow band), and the PDF uncertainties displayed as
a red dashed line show the sensitivity of this measurement on the high $x$ gluon
density specially at high rapidities.

\subsection{Other measurements sensitive to PDFs}
In this section, we describe briefly some other measurements related to QCD and
PDFs measurements.
Another measurement performed by the D\O\ collaboration sensitive to PDFs relies
on the inclusive photon cross section with $|\eta|<0.9$. The measurement is
found to be in good agreement with NLO QCD \cite{photond0} and the extension
of this measurement towards higher rapidity will allow to constrain further the
PDFs. 

The measurement of the inclusive $b$ jet cross sections has been performed by the CDF
collaboration \cite{bjetcdf} using about 300 pb$^{-1}$ and
the comparison with theory suffers large uncertainties related to
renormalisation and factorisation scales. The higher order contributions
which are not included in the theory might play a major role in $b$-jet
production calculations.

Other observables which have been measured by the D\O\ and CDF collaborations
correspond to the jet and $Z$ or $W$ distributions in $Z+$jet or $W+$jet events.
In general, there is a good agreement between the Monte Carlo predictions and
data, but there are some discrepancies in the third jet distributions like for
its transverse momentum for instance \cite{wzd0cdf}. These comparisons are very
important to tune the Monte Carlos for the LHC and the Tevatron since they 
correspond directly to a
background to the search for the Higgs boson.

\section{Diffraction at the Tevatron}
Diffractive events can be analyzed at the Tevatron both in the D\O\ and CDF
experiments. Both experiments installed roman pot detectors in the direction of
the outgoing antiprotons which show a good acceptance down to $t$ close to 0. 
The D\O\ collaboration installed additional roman pot
detectors on both proton and antiproton sides with a good acceptance for $t>0.8$
GeV$^2$ and lower $\xi$. 

In addition to the analyses related to inclusive diffraction measurements using
dijets for instance, the CDF collaboration performed the first search for
exclusive events at the Tevatron. An exclusive production means that the full
available energy is used to produce the object in the main detector (diphoton,
dijets...), or in other words, there is no pomeron remnant.
The existence of the exclusive events is of
special interest for the LHC where the Higgs boson might be produced exclusively
\cite{ushiggs}. The CDF collaboration measured the so-called dijet mass fraction
in dijet events - the ratio of the mass carried by the two jets divided by the
total diffractive mass - when the antiproton is tagged in the roman pot
detectors and when there is a rapidity gap on the proton side to ensure that the
event corresponds to a double pomeron exchange. The results are shown in Fig. 
3 and are compared  with the POMWIG expectation using the gluon and
quark densities measured by the H1 collaboration in dashed line \cite{cdfdiff}. We 
see a clear deficit of
events towards high values of the dijet mass fraction, where exclusive events
are supposed to occur (for exclusive events, the dijet mass fraction is 1 by
definition at generator level and can be smeared out towards lower values taking
into account the detector resolutions). Fig. 3 shows also the
comparison between data and the predictions from the POMWIG and DPEMC
generators,
DPEMC being used to generate exclusive events \cite{ushiggs}. There is
a good agreement between data and MC. However, this does not prove the existence of
exclusive events since the POMWIG prediction shows large uncertainties (the
gluon in the pomeron used in POMWIG is not the latest one shown at this  workshop and 
the
uncertainty at high $\beta$ is quite large \cite{h1gluon}). In addition, it is not 
obvious one
can use the gluon density measured at HERA at the Tevatron since factorisation
is not true, or in other words, this assumes that the survival probability is a
constant, not depending on the kinematics of the interaction.

\begin{wrapfigure}{l}{6.0cm}
\includegraphics[clip,width=6.0cm,height=5.5cm]{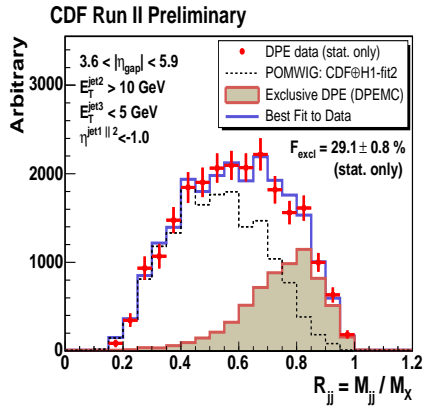}
\caption{Search for exclusive diffractive events at CDF.}
\end{wrapfigure}

The CDF collaboration also looked for the exclusive production of dilepton and
diphoton. Contrary to diphotons, dileptons cannot be produced exclusively via pomeron exchanges since
$g g \rightarrow \gamma \gamma$ is possible, but $g g \rightarrow l^+ l^-$ 
directly is impossible. However, dileptons can be produced via QED processes, and
the cross section is perfectly known. The CDF measurement is $\sigma = 1.6
^{+0.5}_{-0.3} (stat) \pm 0.3 (syst)$ pb which is found to be in good agreement
with QED predictions and shows that the acceptance, efficiencies of the detector
are well understood. 3 exclusive diphoton events have been observed by the CDF
collaboration leading to a cross section of
$\sigma = 0.14
^{+0.14}_{-0.04} (stat) \pm 0.03 (syst)$ pb compatible with the expectations
for exclusive diphoton production at the Tevatron.

\section{$B_S$ oscillation}

\begin{wrapfigure}{l}{5.0cm}
\includegraphics[clip,width=5.0cm,height=4.5cm]{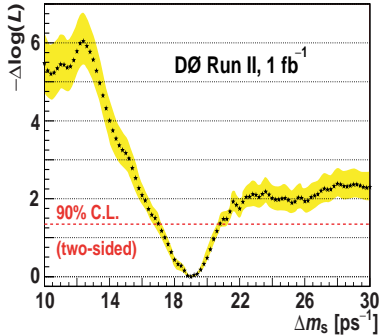}
\caption{$B_S$ oscillation results from D\O\ .}
\end{wrapfigure}

The $B_S$ mesons ($B_S^0$ and $\bar{B_S^0}$) transform via the weak interaction 
between their own constituents
and the rate is characterized by the mass difference between both constituents
called $\Delta m_S$. The standard models predicts that $16.7 < \Delta m_S <25.4$
ps$^{-1}$. For the first time, the D\O\ and CDF experiments have been able to
measure $\Delta m_S$ directly. The D\O\ collaboration looked for events where
$B_S$ decays in $D_S$ mesons ($\bar{c} s$) and a muon and a neutrino. The $D_S$
decays via $\Phi \pi$ to $K^+K^-\pi$. The intersection of the reconstructed
$D_S$ flight path with that of the muon gives the distance the $B_S$ meson
travelled before it decayed, which leads to the lifetime of the $B_S$ meson once
its momentum is determined. The lifetime is then directly related to $\Delta
m_S$. The results are displayed in Fig. 4 which leads to $17 < \Delta m_S < 21$
ps$^{-1}$ at 90\% CL \cite{deltamsd0}. 

The CDF collaboration measured also $\Delta m_S$ more recently with higher
precision benefitting from the layer 0 of their silicon tracker \cite{deltamscdf}. The results are:
$\Delta m_S = 17.33 ^{+0.42} (stat) \pm 0.07 (syst) ps^{-1}$, or
$17.00 <\Delta m_S < 17.91$ ps$^{-1}$ at 90\% CL , 
$16.94 <\Delta m_S < 17.97$ ps$^{-1}$ at 95\% CL 
which leads to $V_{td}/V{ts} = 0.208
^{+0.008}_{-0.007}$.

\section{Electroweak and top quark results}

\subsection{$W$ production}
The $WW$ production cross section was measured recently by the CDF collaboration
using 825 pb$^{-1}$:
 $\sigma  = 
13.6 \pm 2.8 (stat) \pm 1.6 (syst) \pm 1.2 (lum)$ pb, to be compared with the
D\O\ published result with 220-250 pb$^{-1}$,
$\sigma  = 
13.8^{ + 4.3}_{-3.8} (stat) ^{1.2}_{-0.9} (syst) \pm 0.9 (lum)$ pb \cite{wcross}.

Both collaborations measured the $W$ asymmetry \cite{Mikko} which is directly
sensitive to the $d$ over $u$ quark densities.
The measurement stills suffers large statistical uncertainties and will be
improved in a near future. 

The measurement of the $W$ mass is still in progress and should be presented by
Winter 2007. Its precision is expected to be of the order of 20-30 MeV by the
end of Run II.

\subsection{Top quark and electroweak fits}
The top quark production cross section as well as mass measurements are one 
of the main goal of the Tevatron experiments. The measurement of the $t \bar{t}$
production was performed in different multilepton and multijet channels
\cite{crossttbar} and the combined result for the CDF experiment as an example is:
$\sigma = 7.3 \pm 0.5 (stat.) \pm 0.6 (syst.) \pm 0.4 (lum.)$ with 760 pb$^{-1}$. 

A new top mass combination from the D\O\ and CDF collaborations leads
to a much better precision than at Run I (the precision for Run I was 4.4 GeV):
$M_{top}=172.5 \pm 1.3 (stat.) \pm 1.9 (syst.)$ \cite{topmass}. The new top mass
measurement together with new data on $W$ boson width and mass from the Tevatron
and LEP experiments lead to a new Higgs mass from electroweak fits \cite{electroweakfit}:
 $M_{Higgs} = 89 + 42 - 30$ GeV at 68\% CL, and less than 175 GeV at 95\% CL.
The final error on the top mass at the end of Run II at the Tevatron will be of
the order of 1.5 GeV and will constrain further the Higgs boson mass.

The top quark can also produced singly directly via electroweak production
together with a $b$ quark. This process has not yet been observed experimentally
which leads to the following limits: 5.0 pb  ($s$-channel production),
4.4 pb with a luminosity
of 370 pb$^{-1}$  from the D\O\ collaboration, and 
3.2 pb  ($s$-channel production),
3.1 pb ($t$ channel), 3.3 pb (both channels) with a lumi
of 695 pb$^{-1}$ from the CDF collaboration. The SM observation is expected at 3
$\sigma$ with 1.5 fb$^{-1}$ and at 5 $\sigma$ with 4 fb$^{-1}$.

%\begin{figure}[ht]
%\centerline{\epsfxsize=2.3in\epsfbox{combo_xs_750pb.eps}}   
%\caption{$t \bar{t}$ cross section measurements from CDF.}
%\end{figure}

\vspace{-0.1cm}

\section{Search for the Standard Model Higgs boson}

%\begin{wrapfigure}{l}{7.2cm}
%\vspace{-1.8cm}
%\includegraphics[clip,width=7.2cm,height=6.7cm]{pratio_cdfd0_7.eps}
%\caption{Higgs boson search combination plot.}
%\end{wrapfigure}

\begin{figure}[ht]
%\epsfxsize=10cm   %width of figure - will enlarge/reduce the figures
%\epsfbox{fig3.eps}
%\figurebox{2cm}{3cm}{} %to have a box alone 
\vspace{-1cm}
\centerline{\epsfxsize=3.8in\epsfbox{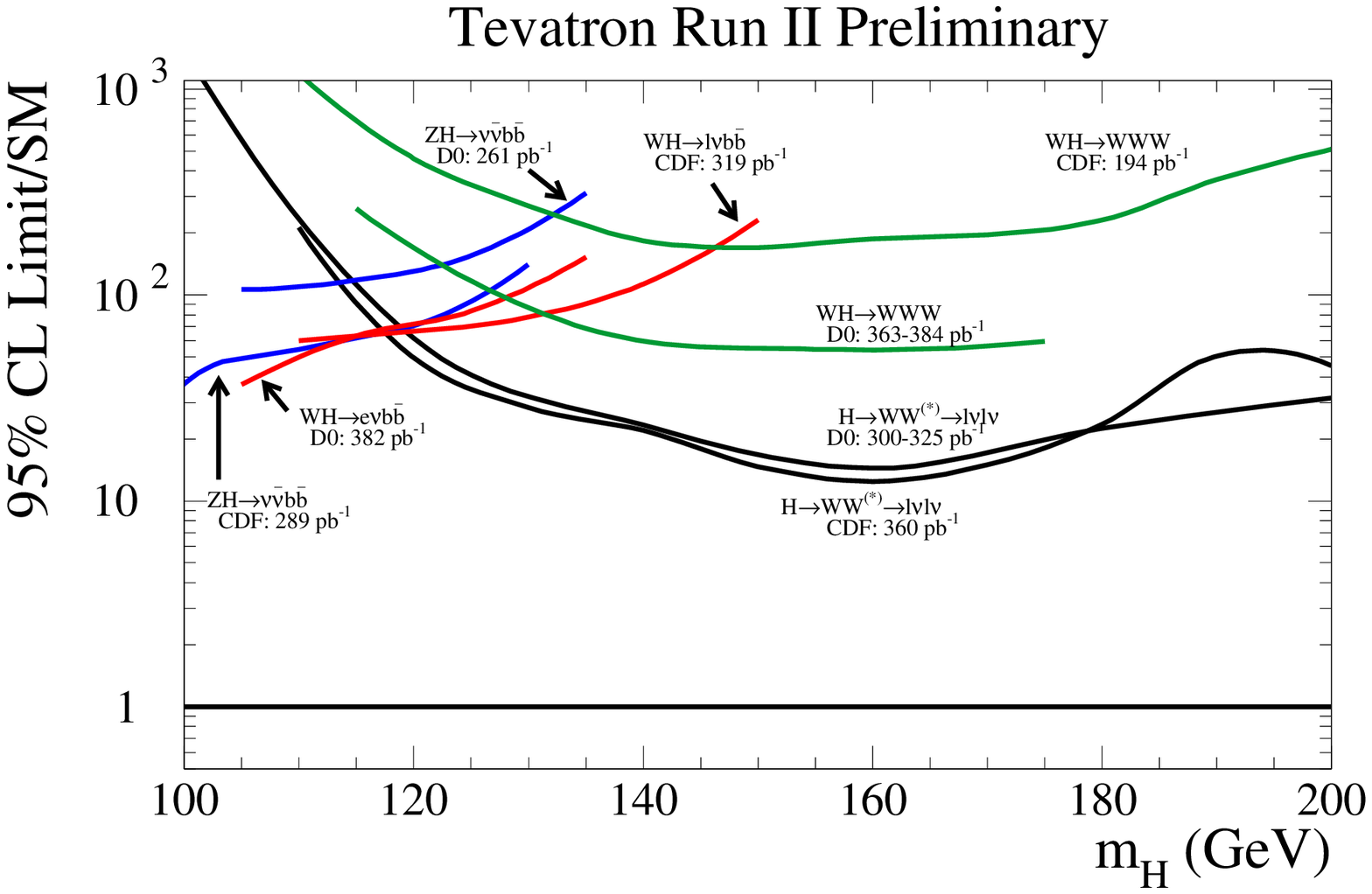}}   
\caption{Higgs boson search combination plot.}
\end{figure}

The search for the SM Higgs boson was performed by both D\O\ and CDF
collaborations \cite{higgs} and the combination is given in Fig. 5. The 95\% CL limit divided
by the SM expectation is displayed in Fig. 5 and shows the present reach of the
Tevatron experiments.

The search for the Higgs boson is made either via
associate production with a $Z$ or a $W$ or via a direct production, and the
Higgs boson either decays in $b \bar{b}$ where the dijet background is quite
large or in $WW$ at higher masses. The present sensitivity of the Tevatron
experiments with less than 400 pb$^{-1}$ is about a factor 10 lower than the SM
expectation for a Higgs boson mass of about 160 GeV. Without any improvement on
the analysis performances, it will be possible to be sensitive on SM Higgs
production at the Tevatron with the full luminosity accumulated by each
experiment of 4-5 fb$^{-1}$ before 2009.

\section{Searches for new phenomena}

\begin{wrapfigure}{l}{6.0cm}
\vspace{-1cm}
\includegraphics[clip,width=6.0cm,height=5.5cm]{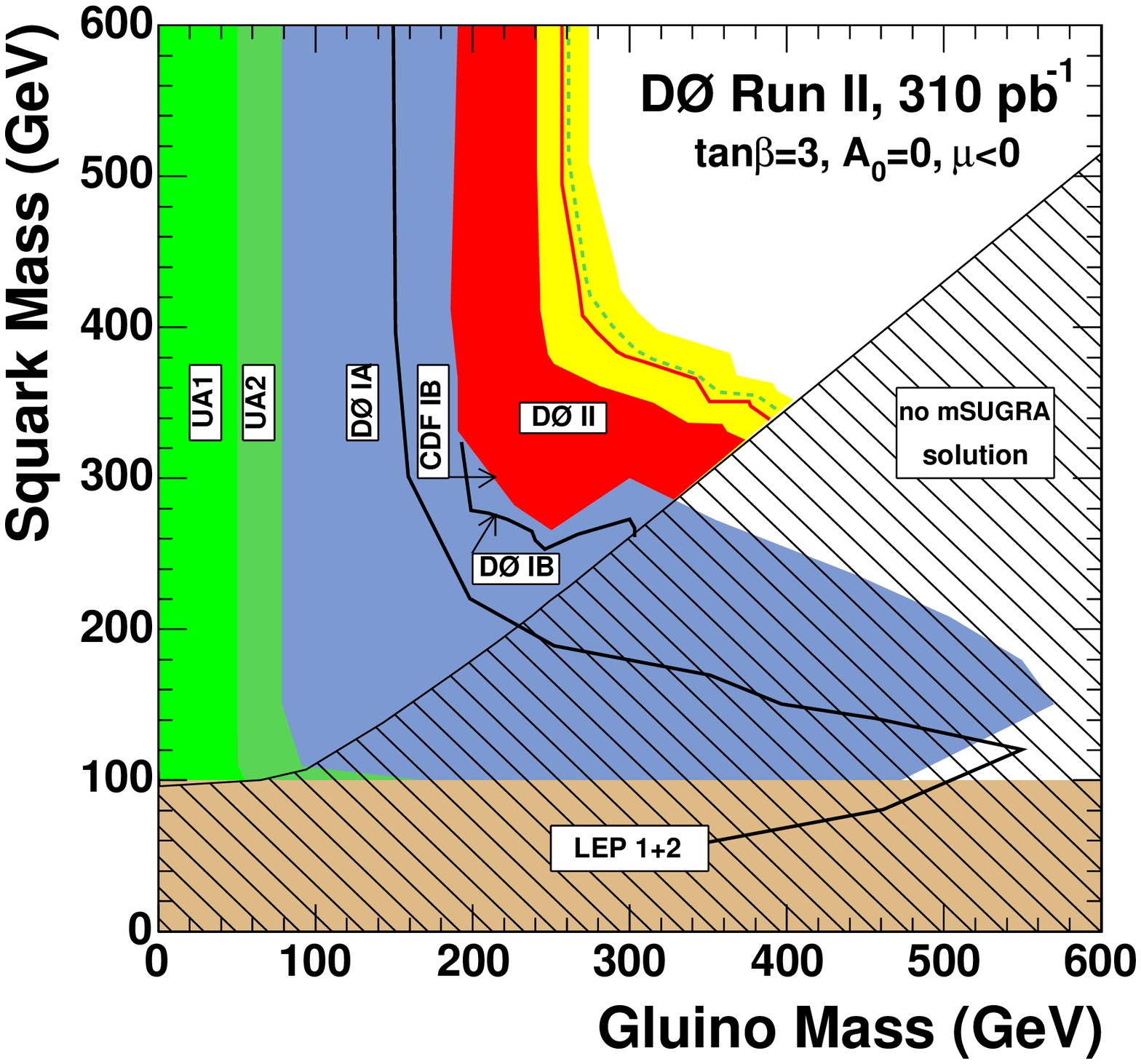}
\caption{Exclusion plot in the gluino and squark mass plane from the D\O\
experiment.}
\end{wrapfigure}

In this report, we will describe only one result concerning the search for
squarks and gluinos in the multijet and missing transverse energy channel
performed by the D\O\ collaboration. Many other searches are performed at the
Tevatron such as stops, leptoquarks, higher dimensions, GMSB and more
information can be found in Ref. \cite{newphenomena}. 

The search for squarks and
gluinos is performed in the $N$ jets and missing transverse energy channel where
$N$ is respectively 2, 3 or 4 when $\tilde{q} \tilde{q}$, $\tilde{q} \tilde{g}$,
$\tilde{g} \tilde{g}$ are produced. The limits obtained in the squark and gluino
mass planes are given in Fig. 6 with a luminosity of about 310 pb$^{-1}$.
Squarks and gluinos with masses respectively below 325 and 241 GeV are excluded
at 95\% CL  for $\tan \beta=3$. Let us notice that one of the limiting factors
entering in this analysis is related to the uncertainty on the high $x$ gluon
density, and this will be even more sensitive at the LHC.
\cite{gluinos}.

\section{Conclusion}
In this talk, we described many results from the Tevatron extending from QCD,
diffraction, electroweak, $b$ physics, top, Higgs and new phenomena. The sensitivity of 
the
Tevatron experiments is the highest until the start of the LHC concerning the
reach on compositeness, SUSY, Higgs boson production. The precision on jet
cross section, $W$ and top quark cross section measurements as well as $b$
physics will increase in the near future. Of special interest before the start
of the LHC is the search for the Higgs boson and SUSY, and the top quark 
properties, as well as the search for exclusive events which is a promising
diffractive channel at the LHC.


\begin{thebibliography}{0}
\bibitem{Mikko} M. Voutilainen, these proceedings; 
\bibitem{nlojet++} Z.~Nagy, Phys.~Rev.~Lett.~{\bf 88}, 122003 (2002).
\bibitem{cteq6} J.~Pumplin {\it et al.},  JHEP~{\bf 0207}, 12 (2002).
\bibitem{kidonakis} N. Kidonakis, J.F. Owens, Phys. Rev. D63, 054019 (2001).
\bibitem{cdfjets} O. Norniella, these proceedings;  CDF Collaboration, 
hep-ex/0512020 and hep-ex/0512062.
\bibitem{photond0} D\O\ Coll., hep-ex/0511054.
\bibitem{bjetcdf} D. Jeans, these proceedings.
\bibitem{wzd0cdf} See
http://www-d0.fnal.gov/Run2Physics/WWW/results/higgs.htm and 
http://www-cdf.fnal.gov /physics/new/qcd/QCD.html
\bibitem{ushiggs}   C.~Royon,
  Mod.\ Phys.\ Lett.\ A {\bf 18}, 2169 (2003) and references therein;
M. Boonekamp, R. Peschanski, C. Royon, Phys. Rev. Lett. {\bf  87 } 
(2001) 
251806 and Nucl. Phys. {\bf B669} (2003) 277;
M. Boonekamp, A. De Roeck, R. Peschanski, C. Royon, Phys. Lett.  {\bf  
B550} (2002) 93;
V.A. Khoze, A.D. Martin, M.G. Ryskin, Eur. Phys. J. {\bf C19} (2001) 477 and
Eur. Phys. J. {\bf C24} (2002) 581.
\bibitem{cdfdiff} M. Gallinaro, these proceedings.
\bibitem{pomwig} B. Cox, J. Forshaw, Comput. Phys. Commun. {\bf 144}
(2002) 104.
\bibitem{h1gluon} P. Newman, these proceedings.
\bibitem{deltamsd0} D\O\ Coll., hep-ex/060404.
\bibitem{deltamscdf} CDF Coll., hep-ex/0606027.
\bibitem{wcross} J. Garcia, these proceedings.
\bibitem{crossttbar} S. Cabrera, these proceedings.
\bibitem{topmass} P. Schieferdecker, these proceddings.
\bibitem{electroweakfit} See http://tevewwg.fnal.gov/top/.
\bibitem{higgs} M. Tomoto, these proceedings
\bibitem{newphenomena} F. Badaud, these proceedings.
\bibitem{gluinos} D\O\ Coll., Phys. Lett. {\bf B 638} (2006) 119.
\end{thebibliography}
\end{document}